\documentclass[preprint,10pt]{sigplanconf}

\usepackage{amsmath}
\usepackage{amsfonts}
\usepackage{stmaryrd}

\newtheorem{theorem}{Theorem}{}
{}
\newtheorem{lemma}{Lemma}{}
{}
\newtheorem{definition}{Definition}{}
\newtheorem{example}{Example}{}

\newcommand{\set}[1]{\{#1\}}

\newcommand{\DFP}{\ensuremath{\mbox{{\sl DFP}}}}

\newcommand{\ra}{\rightarrow}
\newcommand{\Lra}{\Leftrightarrow}

\newcommand{\den}[1]{\ensuremath{\llbracket #1 \rrbracket}}

\begin{document}

\setlength{\pdfpageheight}{\paperheight}
\setlength{\pdfpagewidth}{\paperwidth}

\conferenceinfo{ICFP 2015}{Aug 31 - Sept 2, 2015, Vancouver, BC, Canada} 
\copyrightyear{2015} 


\exclusivelicense                



\title{The lambda mechanism in the lambda calculus and in other calculi}

\authorinfo{M.H. van Emden}
           {Department of Computer Science, University of Victoria}

\maketitle

\begin{abstract}
A comparison of Landin's form of lambda calculus
with Church's shows that, independently of the lambda calculus,
there exists a mechanism for converting functions
with arguments indexed by variables to the usual kind of function
where the arguments are indexed numerically.
We call this the ``lambda mechanism''
and show how it can be used in other calculi.  
In first-order predicate logic it can be used
to define new functions and new predicates
in terms of existing ones.
In a purely imperative programming language
it can be used to provide an Algol-like procedure facility.
\end{abstract}






\section{Introduction}

\paragraph{Lambda in programming languages}

Though ``lambda'' was used to name functions in the first Lisp,
this does not imply that this language conforms to the lambda
calculus:
none of the early Lisps had lexical scoping.
This shows that lambda is a mechanism that
exists independently of lambda calculus.

Let us call ``functional programming'' the use of
a programming language based on lambda calculus.
In 1967 Landin described \cite{landin66},
ISWIM, the first programming language
based on lambda calculus.
PAL \cite{evans68} and POP-2 \cite{brstll68},
which were based on ISWIM, became the first implemented
functional programming languages .
It was followed by many such languages,
of which Scheme, Haskell, and the various
forms of ML are the most widely known in computer science circles.
A more recent phenomenon is that there are widely used languages, such
as JavaScript and Python, of which many users don't even know that
there is a subset allowing functional programming.

The most important lesson of functional programming is that
problems are often more easily solved functionally
rather than imperatively.
Yet at times it seems essential to program imperatively.
Monads are a way to combine functional and imperative programming.
We are interested in other ways of enriching imperative
programming, if not by including lambda calculus,
then perhaps with the lambda mechanism.

%
%
 
\paragraph{The lambda mechanism with predicate calculus}
Floyd's verification method \cite{fld67} leads to
an intricate entanglement of imperative code with logic formulas,
which suggests using logic itself as a programming language.
This is of course what happens in logic programming,
an approach that amounts to the use of a particular theorem
prover as execution mechanism.
Because of its highly specialized choices,
logic programming does not exhaust
the possible uses of logic as a programming language.

Independently of logic programming one sees the following
promising features of first-order predicate logic for use
as a programming language:
\begin{enumerate}
\item
Both functions and predicates.
\item
Simple mathematical semantics that is adaptable to
ontologies familiar to program specifiers.
\item \label{item:func}
Potential for defining new functions in terms of existing ones.
\item \label{item:proc}
Potential for defining new procedures in terms of existing ones.
\end{enumerate}

One of the goals of this paper
is to show how to realize the potential
in items \ref{item:func} and \ref{item:proc}
by means of the lambda mechanism.

\paragraph{The lambda mechanism for Algol-like languages}
Procedure calls in Algol 60 with by-name parameters
have a resemblance to beta reduction in lambda calculus.
This suggests reformulating procedure definition and
procedure call by means of the lambda mechanism.
In this way a transition is made from a purely imperative
language to one that shares features with a
functional programming language.

\section{Notation and terminology}\label{sec:notTerm}
We denote the cardinality of a set $V$ by $|V|$.
For a finite $V$ with $|V| = n$, we freely confuse
the finite cardinals with the corresponding ordinals
and loosely refer to them as ``natural'' numbers.
As a result, locutions such as ``for all $i \in n$'' are common
as abbreviation of ``for all $i \in \{0, \ldots, n-1\}$''.

The set of all functions from set $S$ to set $T$ is denoted
$S \ra T$, so that we may write $f \in (S \ra T)$.
To relieve the overloaded term ``domain'' we call
$S$ the {\em source} and
$T$ the {\em target} of $S \ra T$ and of any $f$ belonging to it.

$|S \ra T| = |T|^{|S|}$. Therefore, when $|S| = 1$
we have that $|S \ra T| = |T|$.

The value of $f$ at $x \in S$ is written as $f(x)$
or as $f_x$.
The composition $h$ of $f \in (S \ra T)$ and $g \in (T \ra U)$
is denoted $g\circ f$ and is the function in $S \ra U$
defined by $x \mapsto g(f(x))$ for all $x \in S$.

Tuples are regarded as functions.
The tuple $t \in (n \ra S)$, with $n$ a natural number,
can be written as $(t_0,\ldots,t_{n-1})$.
The tuple $t$ is said to be ``indexed by'' $n$.
Tuples can also be indexed by other sets.
For example, consider a tuple
$f \in (\{x,y,z\} \ra \{0,1\})$
specified by
$f(x) = 0$,
$f(y) = 1$, and
$f(z) = 0$.
We may use instead
the tabular representation of this tuple:
$f =
\begin{tabular}{c|c|c}
$x$ & $y$ & $z$ \\
\hline
$0$ & $1$ & $0$
\end{tabular}
$.

Consider the set $I \ra D$ of tuples.
Any subset of it is called a {\sl relation},
and its type is $2^{I \ra D}$.
$I$ is the {\sl index set} of the relation.
An $n$-ary relation is a subset of $n \ra D$,
where $n$ is a natural number.
A function $f \in (D \ra D)$ defines the binary relation
$
\{(d, f(d)) \mid d \in D\}.
$
Consider a relation $p$ of type $2\ra D$
such that $(x,y) \in p$ and $(x,y') \in p$
imply that $y=y'$.
Such a relation is called a {\sl partial function}.
``Partial function'' might suggest a special case of ``function'',
but it is the other way around.

\section{The lambda mechanism}

\paragraph{The tuple form of lambda notation}
Landin introduced \cite{landin65} what we shall call the
{\sl tuple form} of lambda notation.
The distinction can be introduced by an example.
Consider the lambda calculus expression
\begin{equation}\label{eq:LCabstraction}
\lambda x_0 \ldots \lambda x_{n-1}\;.\;M
\end{equation}
In lambda calculus abstraction happens one variable at a time;
in this example it is repeated $n$ times.
The $n$ variables are assumed distinct;
assuming otherwise leads to strange phenomena.
For example, according to \cite{hinsel86}, Definition 1.22,
$(\lambda x\lambda x. M) N$
rewrites according to $\beta$-reduction
to $[N/x](\lambda x. M)$,
the result of substituting $N$ for $x$ in $\lambda x. M$.
According to \cite{hinsel86}, case (d) in Definition 1.11 applies.
This case states that $[N/x](\lambda x.P) \equiv \lambda x.P$,
so that $(\lambda x\lambda x. M) N$ $\beta$-reduces to
$\lambda x. M$.

The counterpart of (\ref{eq:LCabstraction}) in 
the tuple form of lambda notation is
\begin{equation}\label{eq:TFLCabstraction}
\lambda (x_0, \ldots, x_{n-1})\;.\;M
\end{equation}
It is a single abstraction on an $n$-tuple of distinct variables.

The application of (\ref{eq:LCabstraction})
in sequence to $M_0,\ldots,M_{n-1}$ is written as
$$
(\ldots((\lambda x_0 \ldots \lambda x_{n-1}\;.\;M)M_0)\ldots) M_{n-1}).
$$
The application of (\ref{eq:TFLCabstraction})
to $(M_0,\ldots,M_{n-1})$ is written as
$$
(\lambda (x_0, \ldots, x_{n-1})\;.\;M)(M_0,\ldots, M_{n-1}).
$$

\paragraph{An example of the lambda mechanism}
Any lambda expression with free variables
can be used to specify a function.
Suppose we are interested in functions over a domain $D$.
The lambda expression $x(yy)$ specifies a function in the sense
that, if domain elements are given as values for $x$ and $y$,
then $x(yy)$ is exactly one element of $D$.
The assignment of domain elements to these variables is
a function of type $\{x,y\} \ra D$.
In this sense $x(yy)$ defines a function
and that function is of type $(\{x,y\} \ra D) \ra D$.
In symbols, $\den{x(yy)} \in ((\{x,y\} \ra D) \ra D)$.
Such a function is called a ``binding'' by
Burstall and Lampson \cite{brsLmp88}.

Suppose now that we want to use $x(yy)$ to define a two-argument
function $f \in (D^2 \ra D)$.
We can't identify $f$ with \den{x(yy)} because they are of
different types.
We need a kind of adapter that converts type
$\{x,y\} \ra D$ to type $D^2 \ra D$.
Such conversions are effected by lambda abstraction in 
the tuple form of lambda calculus.

In this example we have that
$\den{\lambda(x,y).x(yy)} \in (D^2 \ra D)$,
whereas 
$\den{x(yy)} \in ((\{x,y\} \ra D) \ra D)$:
two different expressions have denotations of different types.
In general, for a lambda expression $M$ with set $X$ of free
variables  we have $\den{M} \in ((X \ra D) \ra D)$
and
$\den{\lambda(x_0,\ldots,x_{n-1}).M} \in (D^{|X|} \ra D)$,
assuming that $x_0,\ldots,x_{n-1}$ is one of the $n!$ enumerations
of the $n$ variables in $X$.
The conversion of $\den{M}$ to $\den{\lambda(x_0,\ldots,x_{n-1}).M}$
we call the ``lambda mechanism'',
which is at work in 
the tuple form of lambda calculus.
It converts an expression denoting a function in $(X \ra D) \ra D)$
to one denoting a function in $D^{|X|} \ra D$.

\paragraph{Mathematical formulation of the lambda mechanism}

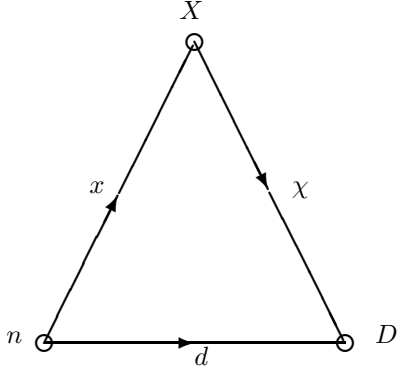
\begin{figure}
\begin{center}
\setlength{\unitlength}{1.0cm}
\begin{picture}(6,6)
\thicklines
\put(1,1){\circle{0.2}}
\put(0.5,1){$n$}
\put(1,1.0){\vector(1,2){0.98}}
\put(2,3.0){\line(1,2){0.98}}
\put(1.6,3.0){$x$}
\put(5,1){\circle{0.2}}
\put(5.4,1){$D$}
\put(1,1){\vector(1,0){2.0}}
\put(3,1){\line(1,0){2.0}}
\put(3.0,0.7){$d$}
\put(3,5){\circle{0.2}}
\put(2.8,5.3){$X$}
\put(3,5.0){\vector(1,-2){0.98}}
\put(4,3.0){\line(1,-2){0.98}}
\put(4.3,3.0){$\chi$}

\end{picture}
\end{center}
\caption{\label{fig:triangle}
$d = \chi \circ x$.
}
\end{figure}

Given an expression $E$ with finite set $X$ of $n$ free variables.
Let $D$, the domain, be a set.
Let $\chi$ be a function in $X \ra D$
(in short, $\chi \in (X \ra D)$),
let $x \in (n \ra X)$, and let $d \in (n \ra D)$
be such that $d(i) = \chi(x(i))$ for all $i \in n$.
See Figure~\ref{fig:triangle}.

The lambda mechanism is the use of $\lambda$ in any
of the following three situations.
\begin{enumerate}
\item\label{item:basic}
Lambda used to transform tuple $\chi$ to tuple $d$:
$d = \lambda x.\chi$.
This is nothing but another way of writing the functional
composition $d =\chi \circ x$.
We come even closer by writing
$$d = x;\chi$$
which is the way some authors
write the functional composition of $x$ and $\chi$.

The fact that in the tuple form of lambda notation
there can be no repeated variables in
$$
x = (x_0,\ldots,x_{n-1})
$$
translates to the existence of the inverse of $x$ regarded as
a function.
Thus we can write $d = \chi \circ x$ equivalently as
$\chi = d \circ x^{-1}$; see footnote\footnote{
It has been suggested \cite{kll15} to write this as
$\chi = \lambda^{-1} x. d$,
but this may be going too far.}
.

\item\label{item:natRel}
A natural extension of a transformation of a tuple to a tuple,
as in item \ref{item:basic}, is to transform a set of tuples
of the same type to a set of tuples of the same type.
That is, a transformation of a relation to a relation when we think of a relation of type $A\ra B$ as a subset of $A\ra B$. 
In our situation, from $P \subseteq (X \ra D)$
to $Q \subseteq (n\ra X)$ so that
$\lambda x. P = Q$ with $Q =\set{\lambda x. p \mid p \in P}$.

This use of lambda is similar to the one found in Section 9.3 of
\cite{lm15}.

\item\label{item:natFunc}
Another natural extension of item \ref{item:basic}
is a transformation from a function $f \in (X\ra D) \ra D$
to a function $g \in (n\ra D)$.
As in item \ref{item:basic},
we have that $d = \lambda x. \chi$.
Suppose that $f$ and $g$ are related by
$f(\chi) = g(d)$ for all $\chi \in (X \ra D)$.
Then we have $g(\lambda x. \chi) = f(\chi)$
for all $\chi \in (X \ra D)$.
As extension to functions of the lambda mechanism,
we write instead $g = \lambda x. f$. 
\end{enumerate}

\section{Predicate logic as programming language}

A functional programming language is one
that is based on the lambda calculus or on combinatory logic.
Similarly, a logic programming language would be one
that is based on predicate logic.
The attraction of the latter is that logic is more popular
for program specification than lambda calculus.

Pure Prolog is a programming language,
and it is based on first-order predicate logic.
It is a special case in several ways:
it is based
on the clausal form of logic,
it presupposes execution by a particular resolution theorem prover,
and its data domain is the Herbrand universe of the
program.
The last restriction is lifted in the Prologs
that are used in practice,
and this has compromised its relation to logic.

The fact that pure Prolog is based on a thin slice of logic
and its tenuous relationship to practice
suggest that we consider anew the potential of first-order predicate
logic as a programming language.

Logic has a lot going for it
as starting point for a programming language.
It has function symbols denoting functions
and predicate symbols denoting relations.
Variable-free terms denote objects
and variable-free formulas denote truth values.
Among the things that are lacking are facilities
to define new functions and relations in terms of existing ones.
In this section we describe how these facilities
can be added by means of the lambda mechanism.

\subsection{Semantics of logic formulas}\label{sec:semForm}

According to an interpretation $I$,
an $n$-ary function symbol $f$ denotes an $n$-argument function
$I(f)$ over a universe of discourse $D$.
An $n$-ary predicate symbol $p$ denotes a relation
$I(p)$ consisting of $n$-tuples of elements of $D$.

\begin{definition}\label{def:meaningVF}
The meaning $M^I$ of variable-free terms and formulas
under interpretation $I$ is defined as follows.
\begin{itemize}
\item
$M^I(c) = I(c)$ if $c$ is a constant.
\item
$M^I(f(t_0,\ldots,t_{n-1}))
=
(I(f))(M^I(t_0),\ldots,M^I(t_{n-1})))$
if $f$ is a function symbol.
\item
$q(t_0,\ldots,t_{k-1})$
is true in I iff
$ (M^I(t_0),\ldots,M^I(t_{k-1}))
   \in I(q)
$
if $q$ is a predicate symbol.
\item
A {\sl conjunction}
$E_0 \wedge \cdots \wedge E_{n-1}$
of formulas
is true in $I$ iff
$E_i$ is true in $I$ for all $i \in n$.
\item
A {\sl disjunction}
$E_0 \vee \cdots \vee E_{n-1}$
of formulas
is true in $I$ iff
$E_i$ is true in $I$ for at least one
$i \in n$.
\item
A formula that is the negation of $E$
is true in $I$ iff $E$ is not true in $I$. 
\end{itemize}
\end{definition}

We regard the formula ``$A$ if $B$'' true if and only if
$A$ is true or the negation of $B$ is true.

We now consider meanings of formulas
with a set $V$ of free variables,
possibly, but not typically, empty.
Let $\alpha$ be an {\sl assignment},
which is a function in $V \rightarrow D$,
assigning an individual in $D$ to every free variable.
In other words, $\alpha$ is a tuple of elements of $D$
indexed by $V$.
As meanings of expressions with variables
depend on $\alpha$, we write $M^I_\alpha$
for the function mapping a term to an element
of the universe $D$
and for mapping a formula to a truth value.

\begin{definition}\label{def:meaningFV}
$M^I_\alpha$ is defined as follows.
\begin{itemize}
\item
$M^I_\alpha(t) = \alpha(t)$ if $t$ is a variable
\item
$M^I_\alpha(c) = I(c)$ if $c$ is a constant
\item
$M^I_\alpha(f(t_0,\ldots,t_{n-1}))
   = (I(f))(M^I_\alpha(t_0),\ldots,M^I_\alpha(t_{n-1}))).
$
\item
$q(t_0,\ldots,t_{k-1})$
is true or false in $I$ with $\alpha$ according to whether
$ (M^I_\alpha(t_0),\ldots,M^I_\alpha(t_{k-1}))
$
is in
$ I(q) $.

\paragraph{}
Now that satisfaction of atoms is defined,
we can continue inductively with satisfaction of
complex formulas.
\item
A conjunction
$E_0 \wedge \cdots \wedge E_{n-1}$
is true in $I$ with $\alpha$
iff $E_i$
is true in $I$ with $\alpha$, for all $i \in n$.
\item
A disjunction
$E_0 \vee \cdots \vee E_{n-1}$
is true in $I$ with $\alpha$
iff $E_i$
is true in $I$ with $\alpha$, for at least one $i \in n$.
\item
If $E$ is a formula,
then $\exists x. E$ is true in $I$ with $\alpha$
iff there is a $d \in D$ such that $E$
is true in
$I$ with $\alpha_{x|d}$
where $\alpha_{x|d}$ is an assignment
that maps $x$ to $d$ and maps the other variables
according to $\alpha$.
\item
If $E$ is a formula,
then $\forall x. E$ is true in $I$ with $\alpha$
iff for all $d \in D$,
$E$ is true in $I$ with $\alpha_{x|d}$
where $\alpha_{x|d}$ is an assignment
that maps $x$ to $d$ and maps the other variables
according to $\alpha$.
\end{itemize}
\end{definition}

\begin{definition}\label{def:bareDenot}
Let $F$ be a formula with set $V$ of free variables.
We define
$$ M^I(F) = \{\alpha \in (V \to D) \mid M_\alpha^I(F)
   \}.
$$
\end{definition}

\subsection{Defining new functions in terms of existing ones}

So far, $I$ has assigned meanings only to variable-free terms.
This is now extended as follows to terms with free variables.

\begin{definition} \label{def:funcSem0}
If $t$ is a term with set $V$ of variables,
then
$M^I(t)$ is the function of type
$(V \to D) \to D$
that maps
$\alpha \in (V \to D)$ to $M^I_\alpha(t) \in D$.
\end{definition}

\begin{example}\label{ex:plusTimes}
$D$ is the set of natural numbers
and $I$ is an interpretation in which
$I(+)$ is addition and $I(\times)$ is multiplication.
$$
(M^I(x+y\times z))(
\begin{tabular}{c|c|c}
$x$ & $y$ & $z$ \\
\hline
$1$ & $2$ & $3$
\end{tabular}
) = 7.
$$
\end{example}
Tabular tuple notation is explained in Section~\ref{sec:notTerm}.

Thus we see that a term $t$
with set $V$ of free variables can be used
to define a function in $(V \to D) \to D$.
However, we cannot use this to define the meaning of a new
function symbol, as this meaning has to be a function
in $(|V| \to D) \to D$.
The required conversion can be made by the lambda mechanism.

\begin{definition}\label{def:funcSem1}
Let $t$ be a term with set $V$ of variables
and let $(x_0,\ldots,x_{n-1})$
be one of the $n!$ enumerations of the $n$ variables in $V$.
Let $f$ be a function symbol that does not occur in $t$
and that is not interpreted by $I$.
We define the result of extending $I$ to be
$I(f) = \lambda(x_0,\ldots,x_{n-1}). M^I(t)$,
where the right-hand side is defined by item~\ref{item:natFunc}
on page~\pageref{item:natFunc}.
\end{definition}

Note that this definition rules out recursivity
both directly and indirectly
via other interpretation extensions.
Recursive definition of a function has to allow
for the possibility that the function is not total.
In first-order predicate logic function symbols denote
total functions.
Partial functions can be defined as binary relations.
We will see that relations can be defined recursively.

\begin{example}
$D$ is the set of natural numbers
and $I$ is an interpretation in which
$I(+)$ is addition and $I(\times)$ is multiplication.
The term $x+y\times z$ of Example~\ref{ex:plusTimes}
can be used to extend $I$ with function symbol $f$
by defining $I(f)$ as $\lambda(x,y,z). M^I(x+y\times z)$
and
by defining $I(g)$ as $\lambda(y,z,x). M^I(x+y\times z)$.
With these extensions of $I$ we have $(M^I(f))(1,2,3) = 7$
and $(M^I(g))(1,2,3) = 5$.
\end{example}

\subsection{Defining new relations in terms of existing ones}
The reason why Definition~\ref{def:funcSem1} rules out recursivity
is that the function symbols denote total functions.
Cartwright \cite{crt84} responds to the need for recursive
definitions by restricting the domain of discourse to
those that are partially ordered with a unique least element
that is interpreted as undefined.

Our response to the need for recursivity
is to represent partial functions via predicate symbols as relations.
In this way there is no need to change the generality of
allowing any domain of discourse, partially ordered or not.
Without changing the classical semantics of
first-order predicate logic, $n$-ary relations of a given type
are partially ordered as sets of tuples and include the empty
tuple.
Because of this the possibility of defining new relations 
from existing ones is more important than that of defining
new functions.
Here also the lambda mechanism is used.

Predicate logic does not provide a facility for
defining new relations.
To make up for this deficiency we introduce
``predicate extensions''.

\paragraph{Syntax of predicate extensions}

\begin{definition}\label{def:predExt}
Let an interpretation $I$ be given.
A {\em predicate extension} is a set of expressions
containing, for $j \in k$ ($k \geq 0$),
$$
p_j := \lambda(x_{j,0},\ldots,x_{j,n_j-1}).
   M^I(F_{j,0} \vee \cdots \vee F_{j,m_j-1})
$$
where each of $F_{j,0}, \ldots, F_{j,m_j-1}$
is a possibly existentially quantified conjunction of atoms.
These expressions satisfy the following two constraints:
(1) no two of the $p_j$ are the same predicate symbol and
(2) $(x_{j,0},\ldots,x_{j,n_j-1})$ is an enumeration of the
free variables in $F_{j,0} \vee \cdots \vee F_{j,m_j-1}$.
\end{definition}

We use ``$:=$'' in the definition of $p_j$ because of the
prevailling convention in logic to use ``$=$'' to denote
the identity relation over the domain of discourse.
See Example~\ref{ex:evenOdd1}.

\begin{example}\label{ex:evenOdd1}
Let an interpretation that $I$ be given that interprets
the function symbols $=$ and $s$.
This interpretation can be extended to one that gives
$even$ and $odd$ mutually recursively defined meanings
as follows.
\begin{eqnarray*}
\{ even &:=&
  \lambda(x). M^I(x = 0 \vee (\exists y. x=s(y) \wedge odd(y))), \\
   odd &:=&
  \lambda(x). M^I(x = s(0) \vee (\exists y. x=s(y) \wedge even(y)))\}
\end{eqnarray*}
\end{example}

\paragraph{Semantics of predicate extensions}

Predicate extensions are syntactic structures that introduce
new symbols.
The intent is to define new relations as denotations of the new
symbols.
As a preparation for such a definition we introduce a class
of interpretations for which the definition is valid.

\begin{definition}
An \DFP-set of interpretations for a given predicate extension
is a set containing interpretations with the following properties:
(1) have the same domain $D$
(2) have the same interpretation for the function symbols of the
predicate extension
(3) have the same interpretation for the predicate symbols that
occur in the right-hand sides and not in the left-hand sides
of the predicate extension.
\end{definition}
Thus the interpretations of a \DFP-set differ only in the
interpretations of the predicate symbols occurring in the left-hand
sides.

\begin{example}\label{ex:evenOdd2}
A \DFP-set for Example~\ref{ex:evenOdd1}
could have the natural numbers as domain $D$,
interpret $s$ as the successor function,
and interpret $=$ as the identity relation over the natural numbers.
\end{example}

\begin{definition}
Given a predicate extension $E$ as in Definition~\ref{def:predExt}.
All \DFP-sets for $E$ are partially ordered by $\preceq$
where $I_0 \preceq I_1$ iff for all predicate symbols $p$
we have $I_0(p) \subseteq I_1(p)$.
The intersection of $I_0$ and $I_1$ is the interpretation $I$
in the \DFP-set for which $I(p) = I_0(p) \cap I_1(p)$
for all predicate symbols $p$ in $E$.
\end{definition}

\begin{definition}\label{def:modelof}
Given a predicate extension $E$ as in Definition~\ref{def:predExt}.
$I$ in a \DFP-set of interpretations for $E$
{\em is a model of} $E$ iff
$$
\lambda(x_{j,0},\ldots,x_{j,n_j-1}).
M^I(F_{j,0} \vee \cdots \vee F_{j,m_j-1}) \subseteq I(p_j)
$$
for all $j \in k$.
\end{definition}

Here $\lambda$ is used according to Item~\ref{item:natRel}
on page~\pageref{item:natRel}
and $M^I$ is used according to Definition~\ref{def:bareDenot}.

\begin{definition}
The {\em formula corresponding to} a predicate extension $P$
as in Definition~\ref{def:predExt} is the formula
$H_0 \wedge \cdots \wedge H_{k-1}$ where, for all $j \in k$,
$H_j$ is
$$
\forall x_{j,0},\ldots,x_{j,n_j-1}.
p_j(x_{j,0},\ldots,x_{j,n_j-1}) \mbox{ if }
(F_{j,0} \vee \cdots \vee F_{j,n_j-1})
$$
\end{definition}

This correspondence allows us
to use some results from \cite{vnmdn14a}.

\begin{lemma}\label{lem:truth2model}
Let $I$ be an interpretation in the \DFP-set of a predicate
extension $P$ as in Definition~\ref{def:predExt}.
The formula corresponding to $P$ is true in $I$ iff
$I$ is a model of $P$ (see Definition~\ref{def:modelof}).
\end{lemma}
{\sl Proof}
Let $\mathcal{F}_j$ abbreviate
$F_{j,0} \vee \cdots \vee F_{j,m_j-1}$.
Let $X_j$ abbreviate $x_{j,0},\ldots,x_{j,n_j-1}$.
For all $j \in k$:
\begin{tabbing}
MMMMMMMMMMMMMMMMMMMMMMMM\= MM\= \kill
$H_j$ is true in $I$ \> $\Lra$ \> (1) \\
$M^I(\mathcal{F}_j)
   \subseteq M^I(p(X_j))$  \> $\Lra$ \> (2) \\
$\lambda(X_j). M^I(\mathcal{F}_j)
   \subseteq \lambda(X_j). M^I(p(X_j))$  \> $\Lra$ \> (3) \\
$\lambda(X_j). M^I(\mathcal{F}_j)
   \subseteq I(p_j)$  \> $\Lra$ \> (4) \\
$I$ is a model of $P$
\end{tabbing}

(1) Lemma 1 in \cite{vnmdn14a},
(2) Monotonicity of $\lambda$,
(3) Definition of $\lambda$ applied to sets of tuples, and
(4) Definition~\ref{def:modelof}.
\hfill $\Box$

\begin{theorem}
Every predicate extension has a minimal model.
\end{theorem}
{\sl Proof}
By Theorem~5 in \cite{vnmdn14a},
every formula corresponding to a predicate extension has a
minimal model.
By Lemma~\ref{lem:truth2model} we conclude that every
predicate extension has a minimal model.
\hfill $\Box$

\begin{example}
With the interpretation of Example~\ref{ex:evenOdd2}
the minimal model assigns to predicate symbol ``even'' (``odd'')
the set of even (odd) numbers.
\end{example}

We conclude that for every predicate extension,
every one of its \DFP-sets has a model that is minimal
in the partial order.
We consider the relations denoted in the minimal model
by the predicates
in the left-hand sides to be the result of the predicate
extension.
In this way we have added to first-order predicate logic
a method for defining new relations in terms of existing ones.
Note its use of the lambda mechanism.

\section{The lambda mechanism for Algol-like languages}
It is desirable in programming
that basic components be easy to write
and that components can be combined with ease
and with few restrictions.
Functional programming is attractive
because functions are such components;
they are easier to combine with fewer restrictions
compared to, say, C.

The unique flexibility of Algol 60 arises from
a number of features (not orthogonal, nor even disjoint):
nested procedure definitions,
procedure calls reminiscent of beta reduction,
lexical scoping,
the call-by-name parameter mechanism.
In the 1960s processor speed was the bottleneck
for all computer applications.
As a result the magic mix of Algol 60 features
was dropped in favour of Pascal and C,
which allowed compilers to generate more efficient code.

In spite of mainstream language and compiler development
going elsewhere,
research into Algol-like languages continued
\cite{reynolds98,hrntnt97,harper12}.
All this work starts with an abstract syntax
and derives semantic equations from it.
Scott domains seem to be necessary even for the
purely imperative subset of the programming languages considered.

Here we are interested in investigating
an alternative approach where the purely imperative subset
is written in Matrix Code \cite{vnmdn14}.
Here the semantics is rigorously defined by
fixpoint methods, but without the mathematical sophistication
of Scott domains.

We take as starting point a purely imperative language;
that is, one without any facility for creating functions
or procedures.
We then add a facility for declaring and calling procedures
modelled on the one of Algol 60.
In the interest of simplicity and clarity
``function procedures'' will not be included
in the experiment.

How to arrive at a minimal imperative language?
Perhaps drop {\sl for}-statements and arrays from Algol 60?
What about switches?
To bypass such questions
we take a radical approach and appeal to the reader's
intuitive understanding of {\sl flowcharts} and exploit the
fact that these can be expressed in Algol 60.
This allows us to replace the considerable amount of detail
that goes into specifying the imperative part of Algol 60
by a compact specification of flowcharts,
which we will leave at an abstract level.

\subsection{Flowcharts without procedures}

\subsubsection{Syntax}\label{subsec:syntax}
``Syntax'' may be a bit misleading,
but it is a useful label to contrast
with Section~\ref{subsec:semantics}, Semantics.
What we are concerned here is {\sl abstract} syntax,
the structure of flowcharts independent of graphical
or textual representation.

A flowchart is a tuple $\langle D,N,B,T \rangle$,
where
\begin{enumerate}
\item
$D$ is a set of declarations.
A declaration allocates a memory location and associates
it with an identifier.
This association is local to the flowchart.
\item
$N$ is a set of {\sl nodes},
\item
$B$ is a set of {\sl boxes}.
A box contains an assignment statement,
which has an identifier as left-hand side
and an arithmetic expression as right-hand side.
\item
$T$ is a set of {\sl tests}.
A test contains a boolean expression.
\end{enumerate}

The components $N$, $B$, and $T$ constitute
the {\sl body} of the flowchart.
The identifiers occurring in the body have
to be declared in $D$.

Boxes, and tests are structured as follows.
A box is a tuple $\langle n_0, a, n_1 \rangle$,
where $n_0$ is a node, the {\sl entry node} of the box,
$a$ is the assignment statement, and
$n_1$ is a node, the {\sl exit node} of the test.

A test is a tuple $\langle n_0, b, n_1, n_2 \rangle$,
where $n_0$ is a node, the {\sl entry node} of the test,
$b$ is the boolean expression,
$n_1$ is a node, the {\sl positive exit node} of the test,
and
$n_2$ is a node, the {\sl negative exit node} of the test.

The nodes are not structured.
They serve to connect boxes and tests
by a node being an exit node of a test or box
and being the entry node of another test or box.
No node can be the entry node of more than one box.
One node, the {\sl start} node, is not an exit node of any box or test.
One node, the {\sl halt} node, is not an entry node of any box or test.

A translation of flowcharts to text can easily be defined.
No new insights will be gained by presenting one here.
We will merely assume that one exists.

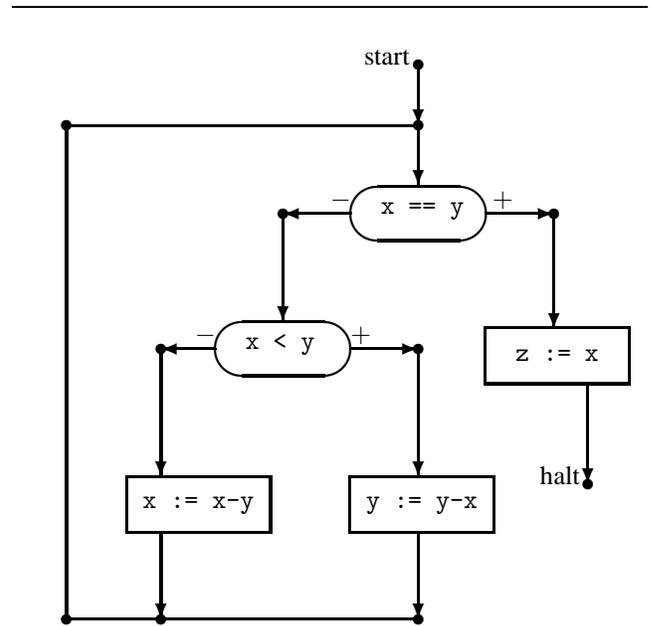
\begin{figure}
\hrule
\setlength{\unitlength}{0.9cm}
\begin{center}
\begin{picture}(10,10.5)
\thicklines
\put(6,9.7){\circle*{0.15}}
\put(5.2,9.7){start}
\put(6,9.7){\vector(0,-1){0.9}}
\put(6,8.8){\circle*{0.15}}
\put(6,8.8){\vector(0,-1){0.9}}
\put(0.8,8.8){\circle*{0.15}}
\put(0.8,8.8){\line(1,0){5.2}}
\put(6.0,7.5){\oval(2,0.8){\hspace{-0.5cm}{\tt x == y}}}
\put(5.0,7.5){\vector(-1,0){1.0}}
\put(4.7,7.6){$-$}
\put(7.0,7.5){\vector(1,0){1.0}}
\put(4.0,7.5){\circle*{0.15}}
\put(4.0,7.5){\vector(0,-1){1.6}}
\put(7.1,7.6){$+$}
\put(8.0,7.5){\circle*{0.15}}
\put(8.0,7.5){\vector(0,-1){1.7}}
\put(4.0,5.5){\oval(2,0.8){\hspace{-0.5cm}{\tt x < y}}}
\put(3.0,5.5){\vector(-1,0){0.8}}
\put(2.7,5.6){$-$}
\put(5.0,5.5){\vector(1,0){1.0}}
\put(5.0,5.6){$+$}
\put(2.2,5.5){\circle*{0.15}}
\put(2.2,5.5){\vector(0,-1){1.9}}
\put(6.0,5.5){\circle*{0.15}}
\put(6.0,5.5){\vector(0,-1){1.9}}
\put(7.0,5.0){\framebox(2.1,0.8){{\tt z := x}}}
\put(8.5,5.0){\vector(0,-1){1.5}}
\put(8.5,3.5){\circle*{0.15}}
\put(7.8,3.5){halt}
\put(1.7,2.8){\framebox(2.1,0.8){{\tt x := x-y}}}
\put(5.0,2.8){\framebox(2.1,0.8){{\tt y := y-x}}}
\put(2.2,2.8){\vector(0,-1){1.3}}
\put(6.0,2.8){\vector(0,-1){1.3}}
\put(2.2,1.5){\circle*{0.15}}
\put(6.0,1.5){\circle*{0.15}}
\put(6.0,1.5){\line(-1,0){3.8}}
\put(0.8,1.5){\circle*{0.15}}
\put(2.2,1.5){\line(-1,0){1.4}}
\put(0.8,1.5){\line(0,1){7.3}}

%

\end{picture}
\end{center}
\caption{\label{fig:flowchExample}
Example of a flowchart.
Every small filled circle represents a node.
Two such circles that are connected
by a line without an arrow
represent the same node.
They are shown separated only for the convenience of 
graphical representation.
Boxes are shown as rectangles; tests as ovals.
A line with an arrow pointing away from a box or test
points to an exit node of the box or test.
A line with an arrow pointing towards a box or test
comes from an entry node of the box or test.
The positive and negative exit nodes of a test
are indicated by plus and minus symbols, respectively.
}
\end{figure}

\subsubsection{Semantics}\label{subsec:semantics}

\paragraph{Operational semantics}
Executing the declarations results
in the creation of the {\sl environment}
which, in the absence of procedures,
consists only of a tuple of locations indexed by indentifiers.

The {\sl state} of a flowchart is
a tuple $\langle k,d \rangle$
where $k$, the {\sl control state}, is a node
and where $d$, the {\sl data state},
is the contents of the tuple
created by executing the declarations.

A {\sl transition} of a flowchart is a change
from state $\langle k,d \rangle$
to state $\langle k',d' \rangle$.
If $k$ is the entry node of a box,
then $k'$ is the exit node of that box
and $d'$ is the result of executing the assignment
statement of the box starting in data state $d$.
If $k$ is the entry node of a test,
then $d' = d$ and
$k'$ is the positive (negative) exit node of that test
if evaluation of the boolean expression in data state $d$
yields true (false).

A state is the {\sl successor of} a state if there is a transition
of the former to the latter.
Every state has a successor except for the states
in which the control state is the halt node\footnote{
The operations in arithmetic and in boolean expressions
are built-in and their executions always terminate.
}.

A {\sl computation} of a flowchart is a sequence of states
in which the first state has the start node as control state
and in which every next state is the successor of the
state preceding it in the computation.

The meaning of a flowchart according to operational semantics
is the binary relation on data states
consisting of all pairs $\langle d,d' \rangle$
such that there is a computation
beginning with $\langle s,d \rangle$
and ending with $\langle h,d' \rangle$,
where $s$ and $h$ are start and halt nodes, respectively. 

\paragraph{Declarative semantics}
The declarative semantics of a box with assignment $a$
is a binary relation of data states:
the set of all $\langle d,d' \rangle$
such that executing $a$ with $d$ as data state results in
data state $d'$.

The declarative semantics of a test with boolean expression $b$
is a pair of complementary subsets of the identity relation on
data states, a positive subset and a negative subset.
The positive (negative) subset is the set of all 
$\langle d,d \rangle$ such that evaluation of $b$ yields
{\sl true} ({\sl false}).

A declarative semantics of a flowchart
can be defined by means of a matrix $M$ of
which the rows and columns are indexed by the nodes of the
flowchart and of which the elements are binary relations over
the data states of the flowchart.
For every box with entry node $i$ and exit node $j$
$M[j,i]$ is the relation denoted by that box.
For every test with entry node $i$,
positive exit node $j$, and
negative exit node $k$,
$M[j,i]$ is the positive and
$M[k,i]$ is the negative 
part of the decomposition of the identity denoted by the test.
All other elements of $M$ are the identity relation over
data states.

The matrix representation of flowcharts shows them to be
an instance of the ``dual-state machines'' of \cite{vnmdn14}.
Theorem 2 of \cite{vnmdn14} implies that the operational
and declarative semantics as given here are equivalent;
that is, define the same binary relation.


\subsection{Flowcharts with procedures}

In Algol 60 procedure calls are reminiscent
of substitution in lambda calculus.
In fact, it is plausible that the lambda calculus and Algol 60
share a common origin in informal mathematics dating back to
at least early 19th century.
Consider for example $f$ defined by
\begin{equation}\label{eq:freeVarj}
f(j) = \Sigma_{i=1}^n ij
\end{equation}
In mathematics it goes without saying that
\begin{itemize}
\item
the value of
(\ref{eq:freeVarj})
is not changed if the occurrences of $i$ are replaced by
any other variable, except j, and
\item
that such a change is mandatory if 
(\ref{eq:freeVarj}) is substituted for $f$
in $\Sigma_{i=1}^{n} f(i)$.
\end{itemize}

\noindent
Similarly we have, in lambda calculus
\begin{quote}
Definition 1.11 ({\sl Substitution})
For any $M$, $N$, $x$ define $[N/x]M$ to be the result
of substituting for every free occurrence of $x$ in $M$,
and changing bound variables to avoid clashes.
\end{quote}
This is from Hindley and Seldin \cite{hinsel86},
where it is followed by a precise definition.

In the definition of Algol 60 \cite{algol60}
we find a similar stipulation:
\begin{quote} \label{quote:4733}
4.7.3.3 Body replacement and execution.
Subsequently the body, modified in this way,
is inserted in place of the procedure statement and is executed.
If the procedure is called from a place
outside the scope of any quantity non-local to the procedure body,
then any conflicts between the identifiers
inserted through this process of body replacement
and the identifiers whose declarations
are valid at the place of the procedure statement
are avoided by suitable systematic changes of the latter identifiers.
\end{quote}

So far boxes in flowcharts can only contain assignment statements.
Let us consider adding the possibility that the state transition
effected by a box is the result of a procedure call
and that the body of the procedure is a flowchart.

%

\subsubsection{Using the lambda mechanism}

The idea behind the lambda mechanism
is that any expression with free variables defines a function.
Here the term ``variable'' derives from lambda calculus.
In the context of flowcharts it is prudent to avoid this term,
so we use ``location'' for what is usually called ``variable''
in imperative languages
and ``formal parameter'' for ``variable'' in the context of
the lambda mechanism.

Thus the idea behind the lambda mechanism
that any expression with free variables defines a function
is rephrased in the context of flowcharts to the idea
that any flowchart where certain identifiers have been
designated as formal parameters defines a function of which
the values are binary relations over states.
Such a function is called ``procedure''.

The most flexible procedure mechanism would allow
any identifier in a statement to be designated as a parameter.
We propose to follow the rule of Algol 60 where this is indeed
allowed and where the only limitation on procedure calls
is that the replacement of actual parameters by formal parameters
has to yield a valid statement.
This rule allows an identifier in the
left-hand side of an assignment statement to be a formal parameter
and forbids the corresponding actual parameter to be
anything but an identifier associated with a location.

\begin{definition}\label{def:procedure}
A {\em procedure}
is a set of declarations of locations
followed by a set of declarations of procedures
followed by the body of the procedure.
The set of declarations of procedures
consists of, for all $j \in k$ with $k \geq 0$,
$$
p_j = \lambda(x_{j,0},\ldots,x_{j,n_j-1}). P_j
$$
satisfying the following constraints:
(1) each of $P_j$ is a procedure,
(2) no two of the identifiers $p_j$ are the same, and
(3) $(x_{j,0},\ldots,x_{j,n_j-1})$ is an enumeration of the
formal parameters in $P_j$.

The body of the procedure is the body of a flowchart,
except that it may contain identifiers that are not 
declared in the procedure's declaration.
\end{definition}

\begin{figure}
\hrule
\setlength{\unitlength}{0.9cm}
\begin{center}
\begin{picture}(10,9)
\thicklines
\put(2,6.2){\circle*{0.15}}
\put(2,6.2){\vector(0,-1){1.0}}
\put(1.2,6.2){start}
\put(1,4.5){\framebox(2,0.7){{\tt x := x-y}}}
\put(2,4.5){\vector(0,-1){1.0}}
\put(2,3.5){\circle*{0.15}}
\put(2,3.5){\vector(0,-1){0.8}}
\put(0.7,2.0){\framebox(2.5,0.7){{\tt gcd2(x,y,z)}}}
\put(2,2.0){\vector(0,-1){1.0}}
\put(2,1.0){\circle*{0.15}}
\put(1.2,1.0){halt}

\put(7,6.2){\circle*{0.15}}
\put(7,6.2){\vector(0,-1){1.0}}
\put(6.2,6.2){start}
\put(6,4.5){\framebox(2,0.7){{\tt y := y-x}}}
\put(7,4.5){\vector(0,-1){1.0}}
\put(7,3.5){\circle*{0.15}}
\put(7,3.5){\vector(0,-1){0.8}}
\put(5.7,2.0){\framebox(2.5,0.7){{\tt gcd2(x,y,z)}}}
\put(7,2.0){\vector(0,-1){1.0}}
\put(7,1.0){\circle*{0.15}}
\put(6.2,1.0){halt}

\end{picture}
\end{center}
\caption{\label{fig:gcd0and1}
The procedures {\tt gcd0} (left) and {\tt gcd1} (right).
Neither has any declarations.
The identifiers
{\tt x},
{\tt y}, and
{\tt z}
are formal parameters.
}
\end{figure}
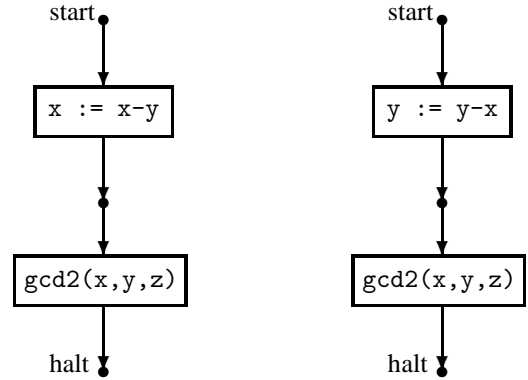

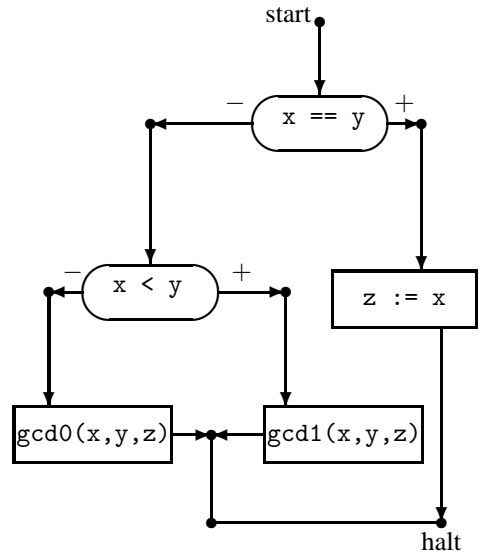
\begin{figure}
\hrule
\setlength{\unitlength}{0.9cm}
\begin{center}
\begin{picture}(10,9)
\thicklines
\put(5,8.5){\circle*{0.15}}
  \put(5,8.5){\vector(0,-1){1.1}}
  \put(4.2,8.5){start}

\put(5,7){\oval(2,0.8){\hspace{-0.5cm}{\tt x == y}}}
  \put(2.5,7){\circle*{0.15}}
  \put(6.5,7){\circle*{0.15}}
  \put(4,7){\vector(-1,0){1.5}}
  \put(3.6,7.2){$-$}
  \put(6,7){\vector(1,0){0.5}}
  \put(6.1,7.2){$+$}

  \put(2.5,7){\vector(0,-1){2.1}}
  \put(6.5,7){\vector(0,-1){2.2}}

\put(2.5,4.5){\oval(2,0.8){\hspace{-0.5cm}{\tt x < y}}}
  \put(1,4.5){\circle*{0.15}}
  \put(4.5,4.5){\circle*{0.15}}
  \put(1.5,4.5){\vector(-1,0){0.5}}
  \put(1.2,4.7){$-$}
  \put(3.5,4.5){\vector(1,0){1.0}}
  \put(3.7,4.7){$+$}

  \put(1,4.5){\vector(0,-1){1.7}}
  \put(4.5,4.5){\vector(0,-1){1.7}}

           \put(5.2,4){\framebox(2.1,0.8){{\tt z := x}}}
           \put(6.8,4){\vector(0,-1){2.9}}
\put(0.5,2){\framebox(2.3,0.8){{\tt gcd0(x,y,z)}}}
  \put(2.8,2.4){\vector(1,0){0.6}}
                 \put(4.2,2){\framebox(2.3,0.8){{\tt gcd1(x,y,z)}}}
                   \put(4.2,2.4){\vector(-1,0){0.8}}
             \put(3.4,2.4){\circle*{0.15}}
             \put(3.4,2.4){\line(0,-1){1.3}}

                  \put(3.4,1.1){\circle*{0.15}}
                          \put(6.8,1.1){\circle*{0.15}}
                          \put(6.5,0.7){halt}
                     \put(3.4,1.1){\line(1,0){3.4}}
\end{picture}
\end{center}
\caption{\label{fig:gcd2}
The procedure {\tt gcd2}.
It has no declarations.
The identifiers
{\tt x},
{\tt y}, and
{\tt z} are parameters.
}
\end{figure}

\begin{example}
This example is a complicated way of computing the
GCD of two numbers. It is distributed over four procedures
with mutual recursion between three of them.
See Figures
\ref{fig:main},
\ref{fig:gcd0and1}, and
\ref{fig:gcd2}.
\end{example}

\begin{figure}
\hrule
\setlength{\unitlength}{1.0cm}
\begin{center}
\begin{picture}(8,10)
\thicklines
\put(1.0,9.5){{\tt nat X,Y,Z;}}
\put(1.0,9.1){{\tt gcd0 = }$\lambda${\tt (x,y,z).G0}}
\put(1.0,8.7){{\tt gcd1 = }$\lambda${\tt (x,y,z).G1}}
\put(1.0,8.3){{\tt gcd2 = }$\lambda${\tt (x,y,z).G2}}
\put(4,7){\circle*{0.15}}
\put(3.2,7){start}
\put(4,7){\vector(0,-1){1.0}}
\put(2.0,5){\framebox(4.0,1.0){{\tt X := 100; Y := 161}}}
\put(4,5){\vector(0,-1){1.0}}
\put(4,4){\circle*{0.15}}
\put(4,4){\vector(0,-1){1.0}}
\put(2.5,2){\framebox(3.0,1.0){{\tt gcd2(X,Y,Z)}}}
\put(4,2){\vector(0,-1){1.0}}
\put(4,1){\circle*{0.15}}
\put(3.2,1){halt}

\end{picture}
\end{center}
\caption{\label{fig:main}
Procedure {\tt main}.
{\tt G0} and {\tt G1}
are the text representations of the flowcharts in
Figure~\ref{fig:gcd0and1}.
{\tt G2}
is the text representation of the flowchart in
Figure~\ref{fig:gcd2}.
}
\end{figure}
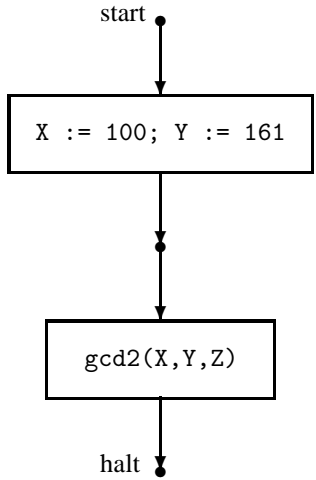

\subsubsection{Operational semantics of the procedure call}
The declarations of the procedure
(see Definition~\ref{def:procedure})
create the environment that determines the effect
of executing the body of the procedure.
This environment consists of two parts.
\begin{enumerate}
\item
A tuple of locations indexed by identifiers.
\item
A tuple of procedures indexed by identifiers.
\end{enumerate}

The effect of the procedure call is specified
when, for every data state of the caller,
it is determined whether the call terminates
and, if so, what the resulting data state of the caller
will be.
This is specified by the following steps.
\begin{enumerate}
\item
Create the environment for the call.
This environment is created by adding the
callee's environment to that of the caller
while omitting declarations in the caller's
environment of variables that are declared in
the callee's environment.
\item
Modify the body of the procedure.
In the body of the procedure formal parameters
are replaced by actual parameters after enclosing
the latter in parentheses wherever syntactically possible.
Possible conflicts between identifiers inserted through
this process and other identifiers already present
within the procedure body are avoided by suitable
systematic changes of the identifiers involved\footnote{
From \cite{algol60}, section 4.7.3.2.
}.
\item
Start execution of the body in the modified environment.
The quote from \cite{algol60} on page~\pageref{quote:4733}
applies.
\item
In case of termination, restore the environment of
the caller on termination.
\end{enumerate}

\section{The lambda mechanism in the lambda calculus}
We have demonstrated the lambda mechanism
in first-order predicate logic
and in an imperative programming language.
In both cases the starting point
was a base language of expressions
of which the meaning could only be determined
with the assignment of values to parameters.
In predicate logic the base language
was that of formulas.
In the imperative programming language
the base language was that of flowcharts.
In both cases the lambda mechanism
made it possible to define procedures,
with partial functions as special case.

With these two examples in front of us,
let us consider the question
whether the lambda calculus
is also an instance of the lambda mechanism.
If so, then there must be a base language.
What is it in the case of the lambda calculus?

The reason why the lambda mechanism
has not been noticed in the lambda calculus
may well be that the base language is so small:
no constants (in the pure lambda calculus,
usually the only form that is studied),
and, apart from one binary operation for application,
only variables.
Moreover, abstraction acts on a single variable.

When we have a lambda expression $N$ with one free variable $x$,
then $N$ by itself denotes a function,
and it is a function of type $(\{x\} \ra D) \ra D$
when we interpret variables as functions of type $D\ra D$.
The difference lies in the distinction in the argument
type $\{x\} \ra D$ in the first case and $D$ in the second case.
As stated in Section~\ref{sec:notTerm},
$|\{x\} \ra D| = D^{|\{x\}|}$.
Thus we see that 
$|\{x\} \ra D| = |D|$.
In other words, there is a bijection between these two sets.
This may explain why we ignore the distinction between them.

\section{Related work}

Predicate extensions are similar to the relational programs
of \cite{vnmdn14a}.
In turn, relational programs are closely related to
Prolog programs.

For the semantics of flowcharts
we have relied on the theory of dual-state automata,
which is the subject of \cite{vnmdn14}.

\section{Future work}
Definition~\ref{def:procedure} for flowcharts with procedures
is similar to
Definition~\ref{def:predExt} for predicate extensions.
Yet for predicate extensions we have given a declarative
semantics, while this is lacking so far for
flowcharts with procedures. 
The reason for the difference is that we followed Algol 60
in allowing formal parameters to be procedure identifiers,
thus making flowcharts with procedures a higher-order
formalism in the sense that predicate extensions stay within
first-order predicate logic. 

If one would disallow procedure identifiers as parameters,
then it seems that one could use predicate extensions as analogy
to define models and to show that a unique minimal model
can be identified as declarative semantics for flowcharts with
procedures.

\section{Conclusions}
Lambda calculus is usually credited to Church's
1941 publication \cite{church41}.
In Church's 1932 paper \cite{church32} he introduces
a lambda {\sl notation}, but not the lambda {\sl calculus}.
In another instance of the distinction,
Landin gives as title of his paper \cite{landin65}
``A correspondence between ALGOL 60 and
Church's Lambda-{\sl notation}''.
This paper proposes a mathematical interpretation
of the distinction.
Because of the added precision
we were emboldened to graduate from mere ``notation''
to the ``lambda mechanism''.

Although the lambda mechanism fails to cover the use of
lambda notation in Church's \cite{church32},
we show that the lambda calculus itself is an instance
and that another instance adds a facility to first-order
predicate logic to define new functions and predicates
in terms of existing ones, thus bringing logic closer to
being usable as a programming language.
Finally, we proposed to base procedures in Algol-like
languages on the lambda mechanism.



\acks

Thanks to Paul McJones for helpful discussions
and valuable information.
I became aware of the distinction between
the variable-indexed and ordinal-indexed versions
of the same relation through discussions with Philip Kelly.

This research benefited from facilities provided
by the University of Victoria and by the Natural Science
and Engineering Research Council of Canada.


\bibliographystyle{abbrvnat}


\end{document}